\documentclass[12pt]{article}

\usepackage[explicit]{titlesec}
\usepackage{hyphenat}
\usepackage[x11names, rgb, table, xcdraw]{xcolor}
\usepackage{tikz}
\usepackage{pdflscape}
\usepackage[T1]{fontenc}
\usepackage{amsmath, amsthm}
\usepackage{graphicx}
\usepackage{hyperref}
\usepackage{spverbatim}
\usepackage{ragged2e} 

\usepackage[paper=portrait,pagesize]{typearea}

\usepackage{soul}
\setul{.6pt}{.4pt}

\usepackage{geometry}
\usepackage{fancyhdr}
\usepackage[labelfont={footnotesize,bf} , textfont=footnotesize]{caption}
\makeatletter
\renewcommand\tagform@[1]{\maketag@@@ {\ignorespaces {\footnotesize{\textbf{Equation}}} #1.\unskip \@@italiccorr }}
\makeatother
\titleformat{\section}
  {\normalfont}{\thesection}{1em}{\MakeUppercase{\textbf{#1}}}
\titlespacing\section{0pt}{0pt}{-10pt}
\titleformat{\subsection}
  {\normalfont}{\thesubsection}{1em}{\textit{#1}}
\titlespacing\subsection{0pt}{0pt}{-8pt}

\makeatletter
\newcommand\sixteen{\@setfontsize\sixteen{16pt}{6}}
\renewcommand{\maketitle}{\bgroup\setlength{\parindent}{0pt}
\begin{flushleft}
\vspace{-.375in}
\sixteen\bfseries \@title
\medskip
\end{flushleft}
\textit{\@author}
\egroup}
\makeatother

\usepackage[sort&compress]{natbib}
\makeatletter
\renewcommand\@biblabel[1]{\textbf{#1.}\hfill}
\makeatother

\bibpunct{}{}{,~}{s}{,}{,}
\title{System Performance with varying L1 Instruction and Data Cache Sizes: An Empirical Analysis}

\author{
        Ramya Akula$^{*a+}$, Kartik Jain$^{*b+}$, and Deep Jigar Kotecha$^{*c+}$\\ \medskip
$^{+}$Computer Science Department, University of Central Florida.\\
$^{a}$\href{mailto:ramya.akula@knights.ucf.edu}{ramya.akula@knights.ucf.edu},\\
$^{b}$\href{mailto:kjain@knights.ucf.edu}{kjain@knights.ucf.edu}, and \\
$^{c}$\href{mailto:deepkotecha@knights.ucf.edu}{deepkotecha@knights.ucf.edu}. \\
$^{*}$Equal Contribution.
}

\begin{document}

\vspace*{.01 in}
\maketitle
\vspace{.12 in}

\section*{abstract}
\justify
In this project, we investigate the fluctuations in performance caused by changing the Instruction (I-cache) size, and the Data (D-cache) size in L1 cache. We employ the Gem5 framework to simulate a system with varying specifications on a single host machine. We utilize the \texttt{FreqMine} benchmark available under the PARSEC suite as the workload program to benchmark our simulated system. The Out-order CPU (O3) with Ruby memory model was simulated in a Full-System X86 environment with Linux OS. The chosen metrics deal with Hit Rate, Misses, Memory Latency, Instruction Rate, and Bus Traffic within the system. Performance observed by varying L1 size within a certain range of values was used to compute Confidence Interval based statistics for relevant metrics. Our expectations, corresponding experimental observations, and discrepancies are also discussed in this report.

\section*{keywords} 
L1 Cache Size; L1 Instruction Cache; L1 Data Cache; Processor Performance; Gem5 Simulator; PARSEC Benchmark - Freqmine. 

\vspace{.12 in}


\section{Introduction}
\justify
Most of the modern fast CPUs poses multiple levels of CPU caches. Earlier CPUs with Cache had only one level of cache, later, level 1 caches is split into L1D for data and L1I for instructions. We believe that it is important to examine the cache size of L1I and L1D due to these reasons: (i)The data that's stored in L1D is different from from L1I. L1I poses not only the instructions, but also annotations such as the next instruction start location, to help out the decoders. "Trace Cache" is used in some processors that stores the result of decoding an instruction rather than storing the original instruction in its encoded form. (ii) Having L1D and L1I separately aids the overall  circuitry, otherwise it would be expensive to have self-modifying code. L1D concerns with read and write operations, while the L1I concerns only the read operation. Here, the write operation goes through L1D to L2 cache, then the line in L1I is invalidated and reloaded from L2. This is an efficient way to access the memory instead of  overwriting the data in L1I directly. (iii) As most of the modern processors can read data from L1D and L1I simultaneously, they suffer from queues at the cache entrance. Having separate caches increases the overall bandwidth such that in a given cycle two reads and one write operations can be performed. (iv)With separate caches, they can power up the circuitry separately for instructions and data, increasing the chances of a circuit remaining un-powered during any given cycle. It is essential to save power; as it is required to the memory cells themselves to maintain their contents, some processors do power down some of the associated circuitry decoders when not in-use. 

\justify
The performance impact of adding an L1 cache is directly related to its efficiency or hit rate, and repeated cache misses can have a catastrophic impact on overall CPU performance.  Since L1 sits at the top of cache hierarchy, we see the problem of minimizing cache misses at the L1 level particularly important for references to levels below and for the execution speed as a whole. Hence, we investigate the effect of changing the size of L1I -L1 Instruction and L1D - L1 Data Caches in full system simulation on an X86 architecture. We aim to use the \texttt{Freqmine} benchmark workload program from the PARSEC suite which fall under the Financial Analysis and the Data Mining application domain respectively. We hypothesize that the fluctuations in the sizes of  L1I -L1 Instruction and L1D - L1 Data caches has a principal effect on the system which rightfully affects the hit rate , the miss rates, and the penalty metrics of the processor. We expect the following changes by changing the L1I and the L1D data caches, the hit rate ideally should increase with the increase in the caches sizes as the hit rate is directly proportional to the cache sizes. Owing to the same logic, we also hypothesize that the miss rate decreases as we increase the size of the caches in the processor system. For a more detailed analysis, we investigated the stats.txt file from the \texttt{Freqmine} workload program benchmark from the PARSEC benchmark suite.

\justify
Following are our system specifications per the project guidelines for achieving full, as well as extra credit:
\begin{itemize}
     \item \textbf{Built and simulated on personal computer} instead of using pre-built libraries on shared system \textit{Eustis}.
     \item \textbf{Full System Linux X86} simulation instead of System-call Emulation.
     \item \textbf{O3 CPU} for out-order, pipelined, and multi-processor execution instead of simpler CPUs.  
     \item \textbf{Ruby} memory model for better flexibility in cache-based systems.
     \item \textbf{Used standardized PARSEC benchmark}.
     \item \textbf{Computed Confidence Interval} based statistics after running a huge number of experiments (\textbf{$>$ 500}).
\end{itemize}

\justify
This report is organized as follows: we mention previous works in the next section. Then we briefly describe introduce the previous two phases: PP1 - Options, PP2 - Experiment Design. Later sections focus on the experiment result statistical analysis and the conclusions drawn from this empirical study. 

\section{Related Work}
\justify
Instruction and data caches are well known architectural solutions that allow significant improvement on the performance of high-end processors. Due to their sensitivity to soft errors they are often disabled in safety critical applications, thus sacrificing performance for improved dependability. The work\cite{2}, reports an accurate analysis of the effects of soft errors in the instruction and data caches of a soft core implementing the SPARC architecture. A cache organization\cite{3} essentially eliminates this penalty. This cache organizational feature has been incorporated in a cache interface subsystem design, and the design has been implemented and prototyped. A master-slave cache system has a large, set-associative master cache, and two smaller direct-mapped slave caches, a slave instruction cache for supplying instructions to an instruction pipeline of a processor, and a slave data cache for supplying data operands to an execution pipeline of the processor. The master cache and the slave caches are tightly coupled to each other. This tight coupling\cite{4} allows the master cache to perform most cache management operations for the slave caches, freeing the slave caches to supply a high bandwidth of instructions and operands to the processor's pipelines. A method\cite{6} for determining a tight bound on the worst case execution time of a program when running on a given hardware system with cache memory. Caches are used to improve the average memory performance, however, their presence complicates the worst case timing analysis. An automatic tool-based approach\cite{5}, to bound worst-case data cache performance. The given approach works on fully optimized code, performs the analysis over the entire control flow of a program, detects and exploits both spatial and temporal locality within data references, produces results typically within a few seconds, and estimates, on average, 30\% tighter WCET bounds than can be predicted without analyzing data cache behavior. According to the study\cite{7}, a cache memory may contain contents that are susceptible to corruption. A cache controller, with the use of a threshold timer, may employ various operations to flush modified cache contents into a main memory and invalidate cache contents so that they are overwritten. Some operations include periodically flushing and invalidating the whole cache memory, periodically flushing and invalidating modified contents, and periodically flushing and invalidating contents based on the time saved in the cache memory. By overwriting cache contents that might otherwise be constantly stored in the cache memory, the system minimizes the probability of cache contents becoming corrupt. The periodic updating of the main memory may also increase the probability of successfully recovering from potential cache parity errors while still maintaining high performance associated with using a cache memory.

\section{Options Selection}
\justify
We follow the steps mentioned in the tutorial\footnote{\href{https://github.com/arm-university/arm-gem5-rsk/wiki}{GitHub Repository for Arm Research Starter Kit}} to run the simulations on gem5. We build a gem5 binary and run a simulation for the X86 processor in Full System(FS) mode \footnote{To avoid redundancy, we mention only the important commands involved as we have already reported detailed installation steps in our project phase 1.}. 

\begin{itemize}
\item We first get and run the clone.sh script to clone both gem5 and arm-gem5-rsk repositories from the aforementioned url.
{\footnotesize
\begin{verbatim}
$ wget https://raw.githubusercontent.com/arm-university/arm-gem5-rsk/master/clone.sh
$ bash clone.sh
\end{verbatim}
}
\item Then, using the tutorial's instructions for ARM, we build gem5 from source but for X86.
{\footnotesize
\begin{verbatim}
$ cd gem5
$ scons build/x86/gem5.opt -j8 # parallel build on 8 host cores.
\end{verbatim}
}
\item We get the X86 full system disk image, expand the image to fit Parsec, and set \$M5\_PATH:
{\footnotesize
\begin{verbatim}
$ wget http://www.m5sim.org/dist/current/x86/x86-system.tar.bz2
$ tar xvfJ x86-system.tar.bz2
\end{verbatim}
}
\item Next, we install and use \texttt{FreqMine} from the PARSEC Benchmark Suite for bench-marking in FS mode for X86.
{\footnotesize
\begin{verbatim}
$ wget http://parsec.cs.princeton.edu/download/3.0/parsec-3.0.tar.gz
$ tar -xvzf parsec-3.0.tar.gz
$ parsecmgmt -a build -c gcc-hooks -p freqmine
\end{verbatim}
}
\item Next, we make some edits to gem5's code for it to work with our setup. More specifcally, we update \texttt{FSConfig.py} so that our \texttt{.img} disk image can be read by gem5.
\item Finally, we run the simulations for all three: \texttt{simsmall}, \texttt{simmedium}, and \texttt{simlarge} as shown below. The exploration parameters are updated for different experimental setups.
{\footnotesize
\begin{verbatim}
$ ~/gem5/build/X86/gem5.opt -d ../../gem5/fs_results/trial_freqmine16x16  ~/gem5/configs
/example/fs.py --disk-image=/home/kartik/gem5/x86_fs_img_files/disks/expanded-linux-x86.img
--kernel=/home/kartik/gem5/x86_fs_img_files/binaries/x86_64-vmlinux-2.6.22.9 --script=/home
/kartik/arm-gem5-rsk/parsec_rcs/freqmine_simsmall_8.rcS --l1i_size="16kB" 
--l1d_size="16kB" --ruby --cpu-type="DerivO3CPU"
\end{verbatim}
}
\end{itemize}

\section{Experiment Design}
\justify
In this section, we briefly describe our experimental design to investigate the processor performance based on varying L1-I and L1-D cache sizes, as proposed in our project phase 2.
\begin{itemize}
    \item \textbf{CPU Parameters:} We used O3 as our pipelined, out-of-order CPU model for the simulation. More specifically, we ran the simulation with Gem5's \textbf{DerivO3CPU} SimObject to simulate this functional unit. We ran experiments with both single, and multiple cores and report our results on the former.
    \item \textbf{Memory Model:} We used \textbf{Ruby} option in the highly configurable \texttt{fs.py} script by Gem5, due to its fidelity and Cache Coherence flexibility. Ruby accurately models both cache coherence and network related features in the memory system. We experiment with L1I, L1D cache sizes and evaluate the performance of the system.
    \item \textbf{Parameter Range for L1 Instruction cache:} L1I cache ranges from  16 KB to 1,024 KB. L1I - Default: 64 KB \& Experimental range: \textbf{[16, 64, 256, 512, 1024] kilobytes}.
    \item \textbf{Parameter Range for L1 Data cache:} L1D cache ranges from  16 KB to 1,024 KB. L1D - Default: 64 KB \& Experimental range: \textbf{[16, 64, 256, 512, 1024] kilobytes}.
    \item \textbf{No of Runs:} For each change in the parameters \textbf{we performed ten runs} to average out any effect of randomness. This gives us a total of 5 parameters for L1I * 5 parameters for L1D * 10 runs for each setup * 3 simulation scales \textit{(small | medium | large)} = \textbf{750 experiments}  to run.
    \item \textbf{Runtime:} The X86 \texttt{Freqmine} simulations were much faster than our initial ARM simulations which allowed us to run a lot of experiments. Because of the minimal boot overhead from the bare-bones Linux X86 image we used, an average simulation took \textbf{45 minutes aprrox.,} to run. We ran multiple such simulations in parallel to make full use of all the cores at our disposal in the host machine.
\end{itemize}

\justify
The number of cores of the simulation subsystem plays a vital role in the estimated run time for each iteration in changes. At our experiment design phase, we made no change to the parameter range for the L1 instruction cache sizes and the L1 data cache sizes. The number of cores we ran the simulation was just one at that moment. Hence, our estimation was based on the stock factory default settings. Therefore, we hypothesized that increasing the number of cores of the subsystem will lead to decrease in the time it requires for the \texttt{Freqmine} benchmark workload program per run.

\section{Experimental Results}

\begin{figure}[h!]
    \centering
    \includegraphics[width=\linewidth]{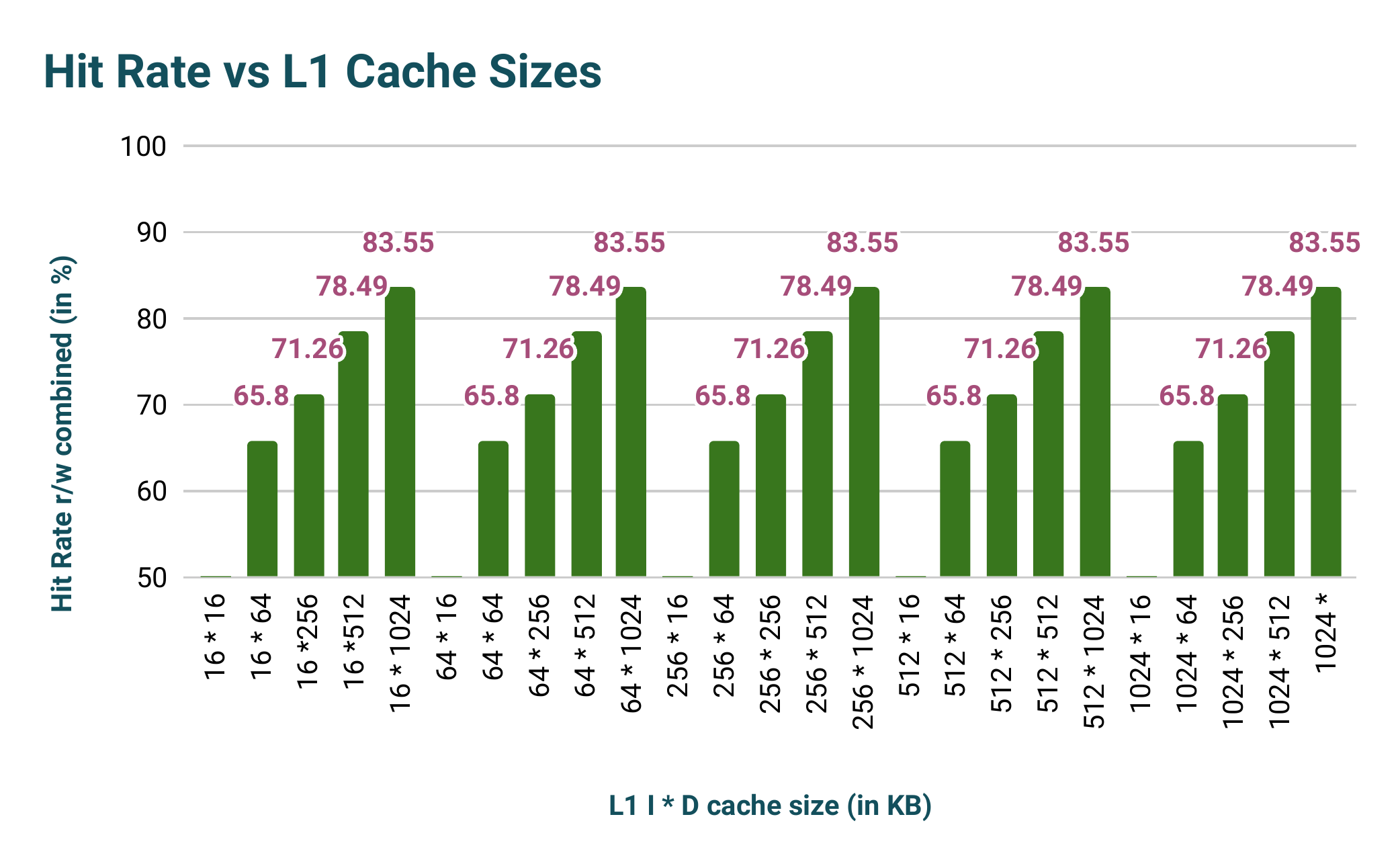}
    \caption{Page hit rate (in percentage) for read and write combined w.r.t. changing I-cache and D-cache sizes in L1. Hit rate is observed to increase proportional to D-cache size. For simulations with extremely small D-cache, the X86 system crashes during benchmarking i.e. \texttt{Freqmine}.}
    \label{fig:hit_rate}
\end{figure}

\begin{figure}[h!]
    \centering
    \includegraphics[width=\linewidth]{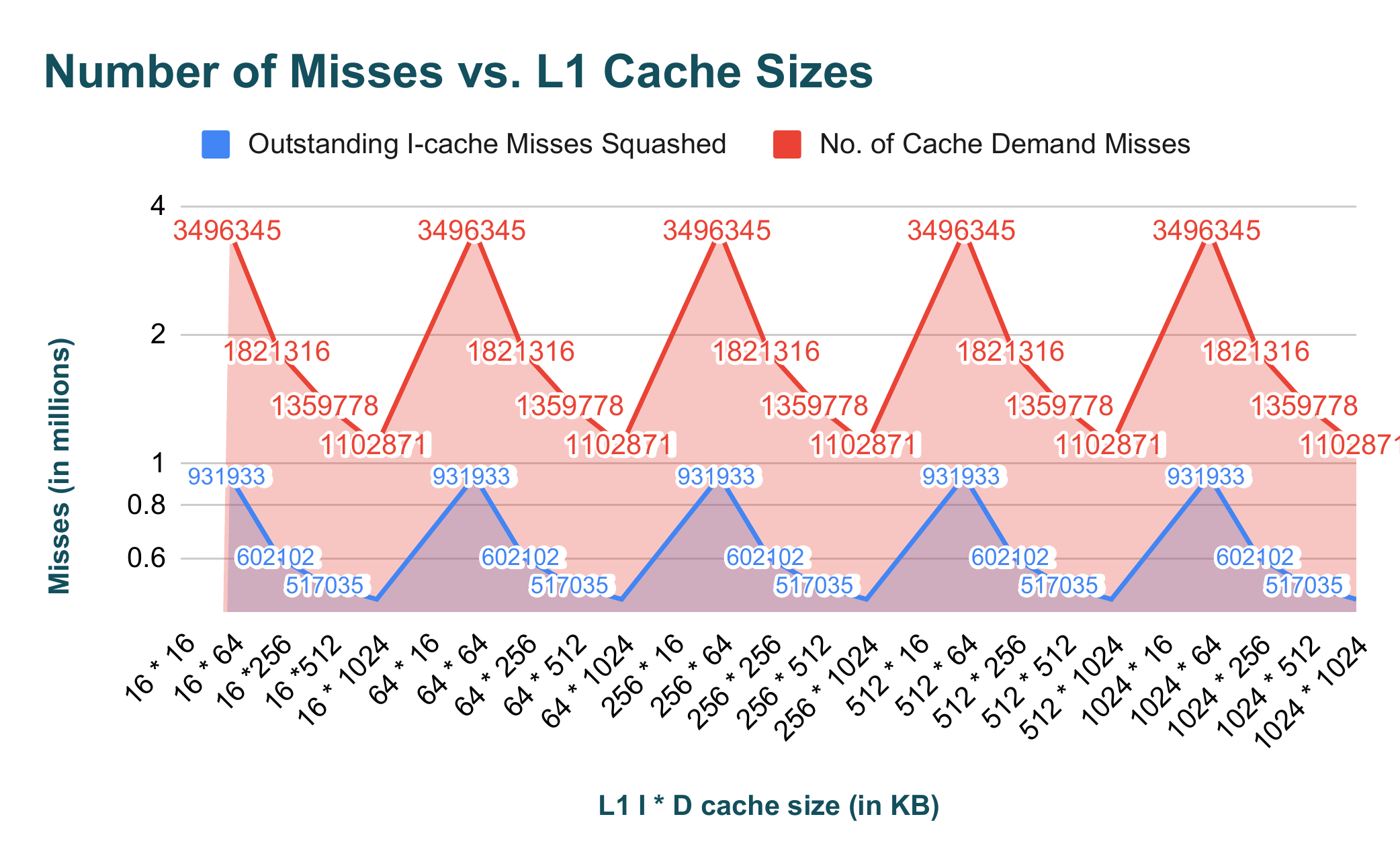}
    \caption{Number of misses (in millions) for \texttt{simmedium} simulation on a log scale w.r.t. changing I-cache and D-cache sizes in L1. Both demand misses, and squashed, outstanding misses are observed to have inverse relation with D-cache size. For simulations with extremely small D-cache, the X86 system crashes during benchmarking i.e. \texttt{Freqmine}.}
    \label{fig:no_of_misses}
\end{figure}

\begin{figure}[h!]
    \centering
    \includegraphics[width=\linewidth]{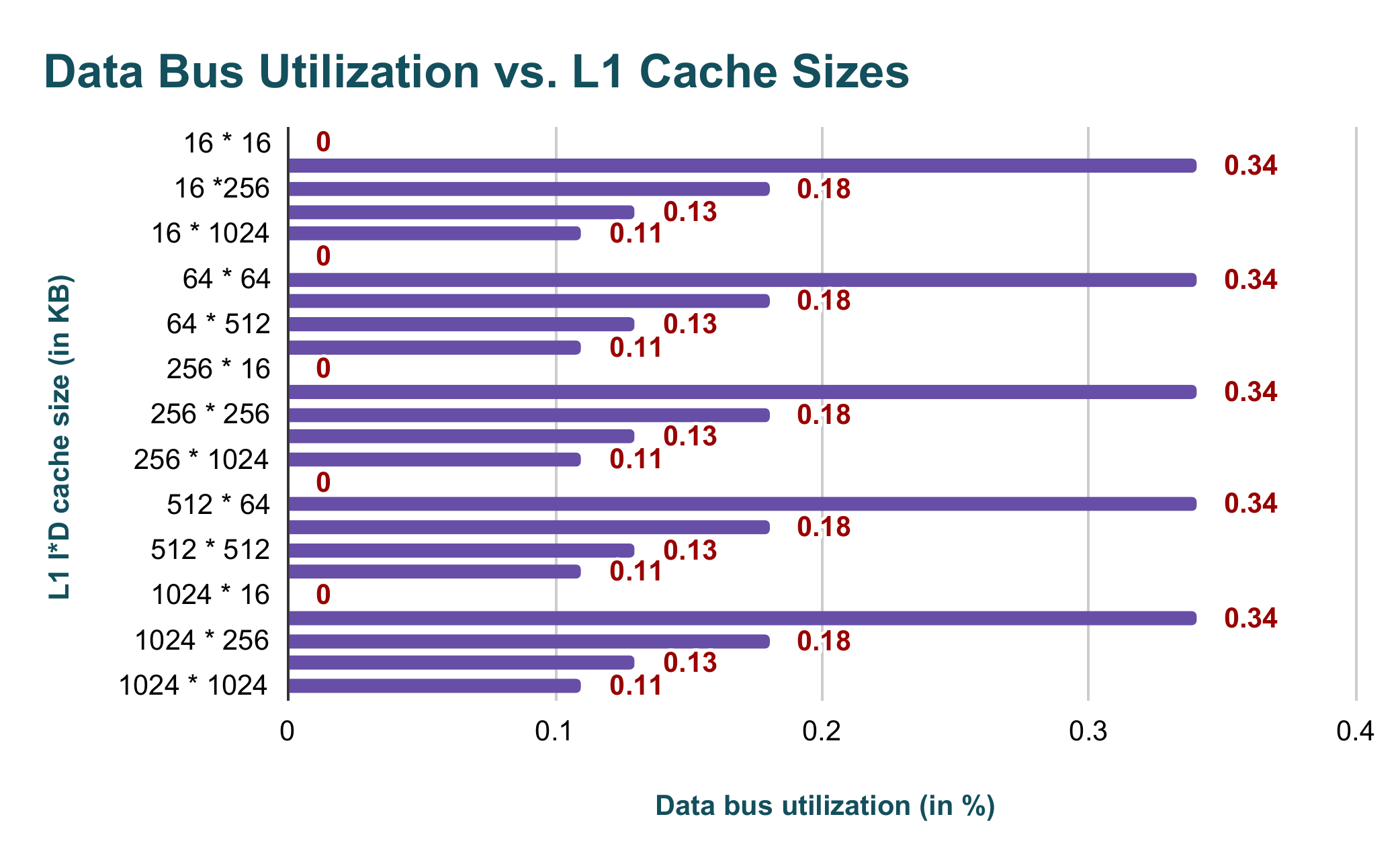}
    \caption{Data bus utilization (in percentage) w.r.t. changing I-cache and D-cache sizes in L1. The utilization is observed to have inverse relation with D-cache size. This implies lower bus traffic for bigger D-caches. For simulations with extremely small D-cache, the X86 system crashes during benchmarking i.e. \texttt{Freqmine}.}
    \label{fig:data_bus_util}
\end{figure}

\begin{figure}[h!]
    \centering
    \includegraphics[width=\linewidth]{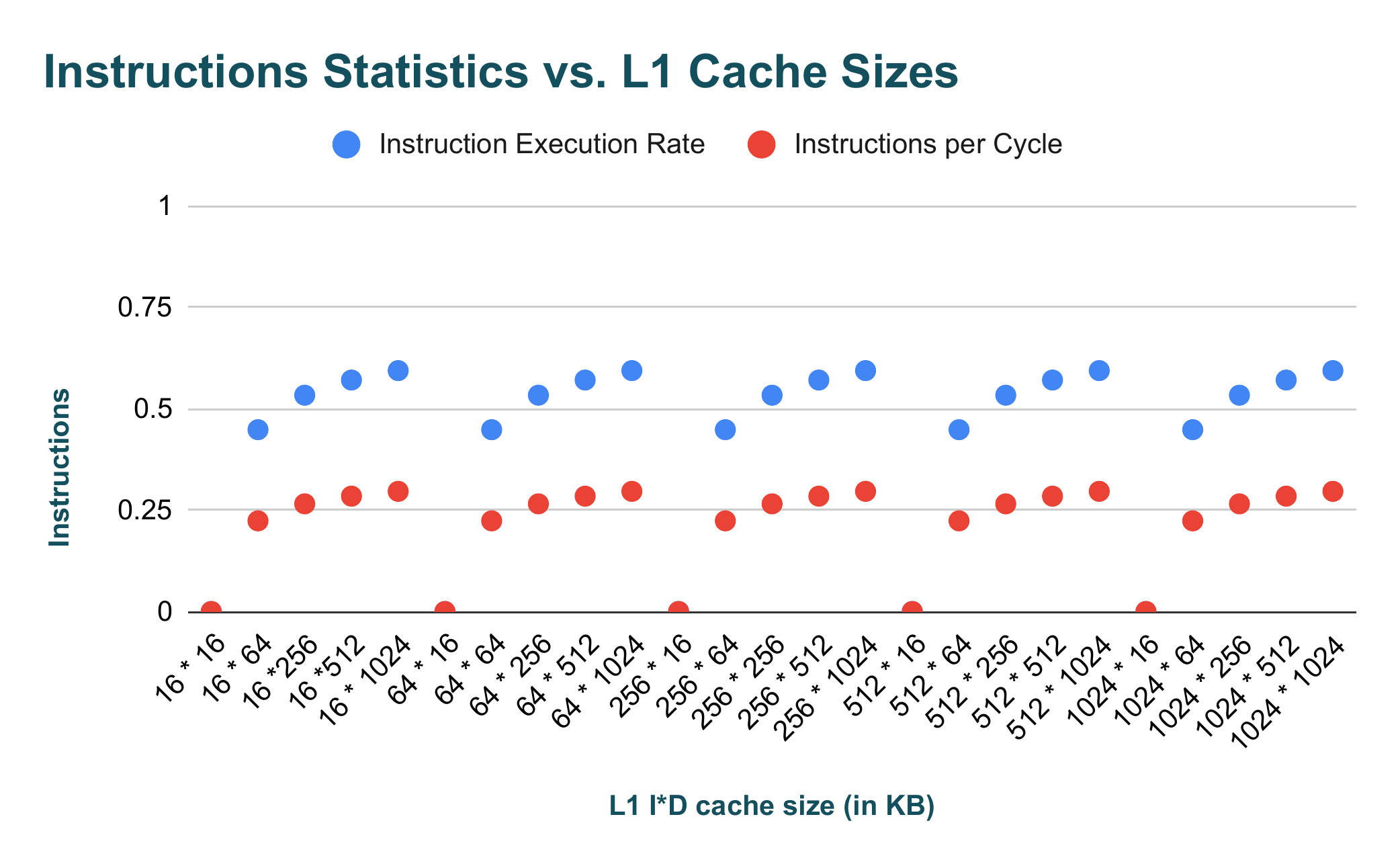}
    \caption{Instruction Execution Rate (IER) and Instructions Per Cycle (IPC) w.r.t. changing I-cache and D-cache sizes in L1. We observe a general trend of higher "speed" for instructions with increase in D-cache size. For simulations with extremely small D-cache, the X86 system crashes during benchmarking i.e. \texttt{Freqmine}.}
    \label{fig:inst_stats}
\end{figure}

\begin{figure}[h!]
    \centering
    \includegraphics[width=\linewidth]{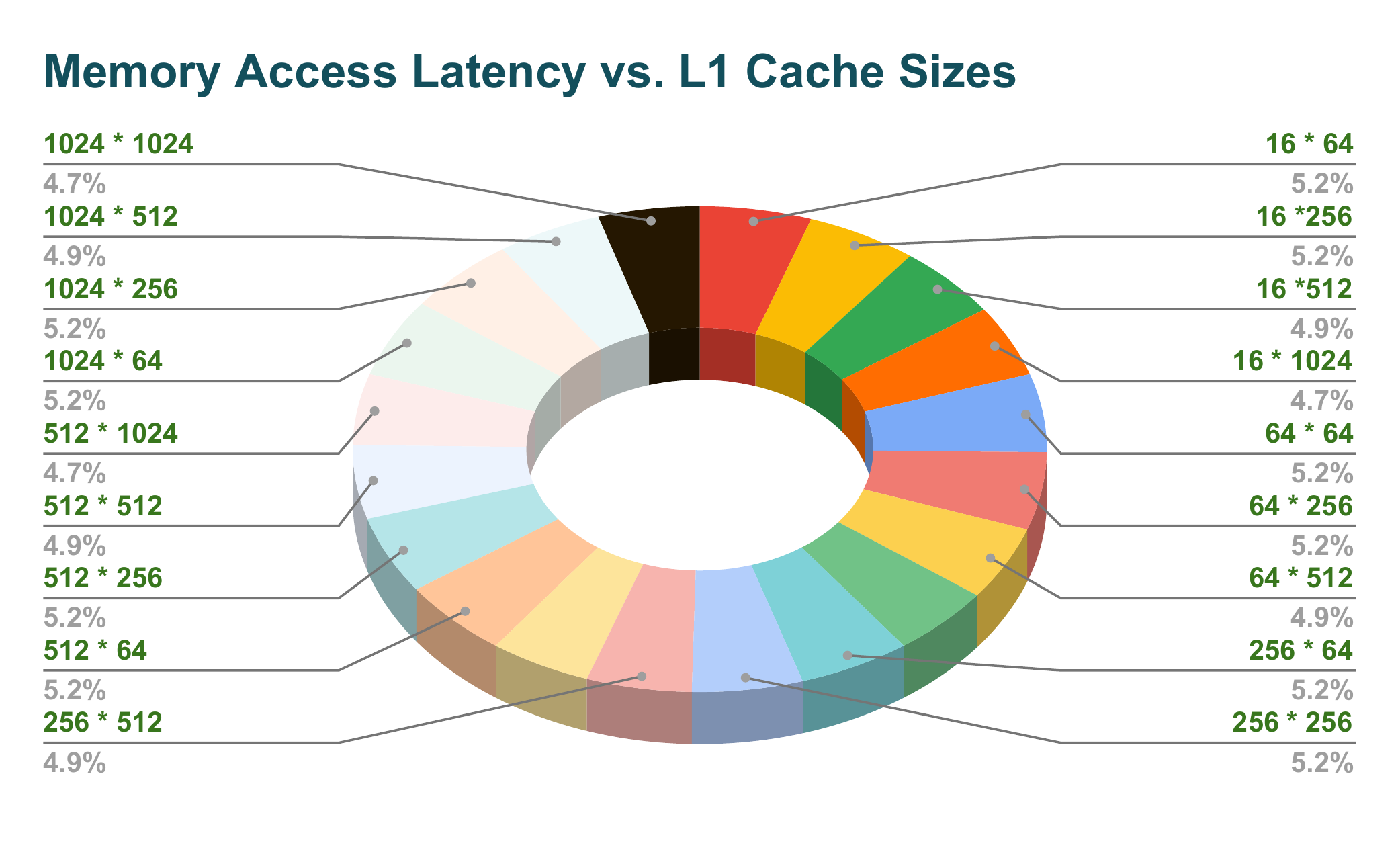}
    \caption{Memory Access Latency (in cycles) per DRAM burst w.r.t. changing I-cache and D-cache sizes in L1. For bigger D-cache, even though more data is moved from memory, fewer access, on average, lead to a shorter latency. For simulations with extremely small D-cache, the X86 system crashes during benchmarking i.e. \texttt{Freqmine}.}
    \label{fig:mem_lat}
\end{figure}

\justify
Based on our experimental design, we performed experiments for 25 (5 unique L1D caches sizes and 5 unique L1I caches sizes) * 3 (simulation size - large, medium, and small) in total. Also, the total number of runs for experiments in total is greater than 500. We used an X86 image which takes about 45 minutes approximately for each experiment. Simulation results across varying cache sizes are shown in Figures [\ref{fig:hit_rate}, \ref{fig:no_of_misses}, \ref{fig:data_bus_util}, \ref{fig:inst_stats}, and \ref{fig:mem_lat}].

We also extract metrics from the \verb+stats.txt+ file for a each experiment using a custom written python program. The custom python program parses the root directories of the gem5, sub-directories of gem5, and crawls to search \verb+stats.txt+ then extract the desired metrics and plots them appropriately. We also verify our simulated configurations through the \verb+config.ini+ files. The scripts and results are zipped and uploaded as a submission comment.

\section{Quantitative Analysis}

\justify
The tables [\ref{tab:large}, \ref{tab:medium}, \ref{tab:small}] represents processor performance across various metrics and runs for large, medium and small simulation sizes respectively. Description for the reported metrics in the tables[\ref{tab:large},  \ref{tab:medium}, \ref{tab:small}] are as follows: 
\begin{itemize}
    \item IPC - Instruction per cycle
    \item IER - Instruction Execution Rate
    \item DBU - Data Bus Utilization
    \item Hit Rate - Hit Rate for read-write combined
    \item Squashed missed - Outstanding I-cache Misses Squashed
    \item \textbf{\#} Cache Misses - Number of cache demand misses
    \item Avg. Mem. per DRAM - Average Memory Access Latency per DRAM burst.
\end{itemize}

\justify
\newgeometry{margin=2cm} 
\begin{landscape}
\begin{table}[] 
\scriptsize
\begin{tabular}{|c|c|c|c|c|c|c|c|c|c|c|c|c|}
\hline
\textbf{Benchmark} & \textbf{Size} & \textbf{L1I Size} & \textbf{L1D size} & \textbf{L1I * L1D} & \textbf{\# Runs} & \textbf{Squashed misses} & \textbf{IER} & \textbf{IPC} & \textbf{\# Cache Misses} & \textbf{Avg. Mem. per DRAM} & \textbf{DBU} & \textbf{Hit rate} \\ \hline
Freqmine & Large & 16 & 16 & 16 * 16 & 10 & 0 & 0 & 0 & 0 & 0 & 0 & 0 \\ \hline
Freqmine & Large & 16 & 64 & 16 * 64 & 1 & 932167 & 0.448235 & 0.223482 & 3496299 & 40550.07 & 0.34 & 65.8 \\ \hline
Freqmine & Large & 16 & 256 & 16 *256 & 1 & 601880 & 0.533339 & 0.26546 & 1820694 & 40206.53 & 0.18 & 71.28 \\ \hline
Freqmine & Large & 16 & 512 & 16 *512 & 1 & 517037 & 0.570672 & 0.284355 & 1359632 & 38216.32 & 0.13 & 78.49 \\ \hline
Freqmine & Large & 16 & 1024 & 16 * 1024 & 1 & 480141 & 0.59391 & 0.296251 & 1102909 & 36830.53 & 0.11 & 83.56 \\ \hline
Freqmine & Large & 64 & 16 & 64 * 16 & 1 & 0 & 0 & 0 & 0 & 0 & 0 & 0 \\ \hline
Freqmine & Large & 64 & 64 & 64 * 64 & 10 & 932167 & 0.448235 & 0.223482 & 3496299 & 40550.07 & 0.34 & 65.8 \\ \hline
Freqmine & Large & 64 & 256 & 64 * 256 & 1 & 601880 & 0.533339 & 0.26546 & 1820694 & 40206.53 & 0.18 & 71.28 \\ \hline
Freqmine & Large & 64 & 512 & 64 * 512 & 1 & 517037 & 0.570672 & 0.284355 & 1359632 & 38216.32 & 0.13 & 78.49 \\ \hline
Freqmine & Large & 64 & 1024 & 64 * 1024 & 1 & 480141 & 0.59391 & 0.296251 & 1102909 & 36830.53 & 0.11 & 83.56 \\ \hline
Freqmine & Large & 256 & 16 & 256 * 16 & 1 & 0 & 0 & 0 & 0 & 0 & 0 & 0 \\ \hline
Freqmine & Large & 256 & 64 & 256 * 64 & 1 & 932167 & 0.448235 & 0.223482 & 3496299 & 40550.07 & 0.34 & 65.8 \\ \hline
Freqmine & Large & 256 & 256 & 256 * 256 & 10 & 601880 & 0.533339 & 0.26546 & 1820694 & 40206.53 & 0.18 & 71.28 \\ \hline
Freqmine & Large & 256 & 512 & 256 * 512 & 1 & 517037 & 0.570672 & 0.284355 & 1359632 & 38216.32 & 0.13 & 78.49 \\ \hline
Freqmine & Large & 256 & 1024 & 256 * 1024 & 1 & 480141 & 0.59391 & 0.296251 & 1359778 & 38220.44 & 0.13 & 78.49 \\ \hline
Freqmine & Large & 512 & 16 & 512 * 16 & 1 & 0 & 0 & 0 & 0 & 0 & 0 & 0 \\ \hline
Freqmine & Large & 512 & 64 & 512 * 64 & 1 & 932167 & 0.448235 & 0.223482 & 3496299 & 40550.07 & 0.34 & 65.8 \\ \hline
Freqmine & Large & 512 & 256 & 512 * 256 & 1 & 601880 & 0.533339 & 0.26546 & 1820694 & 40206.53 & 0.18 & 71.28 \\ \hline
Freqmine & Large & 512 & 512 & 512 * 512 & 10 & 517037 & 0.570672 & 0.284355 & 1359632 & 38216.32 & 0.13 & 78.49 \\ \hline
Freqmine & Large & 512 & 1024 & 512 * 1024 & 1 & 480141 & 0.59391 & 0.296251 & 1102909 & 36830.53 & 0.11 & 83.56 \\ \hline
Freqmine & Large & 1024 & 16 & 1024 * 16 & 1 & 0 & 0 & 0 & 0 & 0 & 0 & 0 \\ \hline
Freqmine & Large & 1024 & 64 & 1024 * 64 & 1 & 932167 & 0.448235 & 0.223482 & 3496299 & 40550.07 & 0.34 & 65.8 \\ \hline
Freqmine & Large & 1024 & 256 & 1024 * 256 & 1 & 601880 & 0.533339 & 0.26546 & 1820694 & 40206.53 & 0.18 & 71.28 \\ \hline
Freqmine & Large & 1024 & 512 & 1024 * 512 & 1 & 517037 & 0.570672 & 0.284355 & 1359632 & 38216.32 & 0.13 & 78.49 \\ \hline
Freqmine & Large & 1024 & 1024 & 1024 * 1024 & 10 & 480141 & 0.59391 & 0.296251 & 1102909 & 36830.53 & 0.11 & 83.56 \\ \hline
\end{tabular}
\caption{\texttt{Freqmine} benchmarks for large simulation on the X86 simulated processor using gem5 simulator.}
\label{tab:large}
\end{table}
\end{landscape}
\newgeometry{margin=2cm} 
\begin{landscape}
\begin{table}[] 
\scriptsize
\begin{tabular}{|c|c|c|c|c|c|c|c|c|c|c|c|c|}
\hline

\textbf{Benchmark} & \textbf{Size} & \textbf{L1I Size} & \textbf{L1D size} & \textbf{L1I * L1D} & \textbf{\# Runs} & \textbf{Squashed misses} & \textbf{IER} & \textbf{IPC} & \textbf{\# Cache Misses} & \textbf{Avg. Mem. per DRAM} & \textbf{DBU} & \textbf{Hit rate} \\ \hline
Freqmine & Medium & 16 & 16 & 16 * 16 & 10 & 0 & 0 & 0 & 0 & 0 & 0 & 0 \\ \hline
Freqmine & Medium & 16 & 64 & 16 * 64 & 1 & 931933 & 0.44823 & 0.223479 & 3496345 & 40548.71 & 0.34 & 65.8 \\ \hline
Freqmine & Medium & 16 & 256 & 16 *256 & 1 & 602102 & 0.533319 & 0.265416 & 1821316 & 40220.09 & 0.18 & 71.26 \\ \hline
Freqmine & Medium & 16 & 512 & 16 *512 & 1 & 517035 & 0.570667 & 0.28436 & 1359778 & 38220.44 & 0.13 & 78.49 \\ \hline
Freqmine & Medium & 16 & 1024 & 16 * 1024 & 1 & 480607 & 0.593974 & 0.296231 & 1102871 & 36834.71 & 0.11 & 83.55 \\ \hline
Freqmine & Medium & 64 & 16 & 64 * 16 & 1 & 0 & 0 & 0 & 0 & 0 & 0 & 0 \\ \hline
Freqmine & Medium & 64 & 64 & 64 * 64 & 10 & 931933 & 0.44823 & 0.223479 & 3496345 & 40548.71 & 0.34 & 65.8 \\ \hline
Freqmine & Medium & 64 & 256 & 64 * 256 & 1 & 602102 & 0.533319 & 0.265416 & 1821316 & 40220.09 & 0.18 & 71.26 \\ \hline
Freqmine & Medium & 64 & 512 & 64 * 512 & 1 & 517035 & 0.570667 & 0.28436 & 1359778 & 38220.44 & 0.13 & 78.49 \\ \hline
Freqmine & Medium & 64 & 1024 & 64 * 1024 & 1 & 480607 & 0.593974 & 0.296231 & 1102871 & 36834.71 & 0.11 & 83.55 \\ \hline
Freqmine & Medium & 256 & 16 & 256 * 16 & 1 & 0 & 0 & 0 & 0 & 0 & 0 & 0 \\ \hline
Freqmine & Medium & 256 & 64 & 256 * 64 & 1 & 931933 & 0.44823 & 0.223479 & 3496345 & 40548.71 & 0.34 & 65.8 \\ \hline
Freqmine & Medium & 256 & 256 & 256 * 256 & 10 & 602102 & 0.533319 & 0.265416 & 1821316 & 40220.09 & 0.18 & 71.26 \\ \hline
Freqmine & Medium & 256 & 512 & 256 * 512 & 1 & 517035 & 0.570667 & 0.28436 & 1359778 & 38220.44 & 0.13 & 78.49 \\ \hline
Freqmine & Medium & 256 & 1024 & 256 * 1024 & 1 & 480607 & 0.593974 & 0.296231 & 1102871 & 36834.71 & 0.11 & 83.55 \\ \hline
Freqmine & Medium & 512 & 16 & 512 * 16 & 1 & 0 & 0 & 0 & 0 & 0 & 0 & 0 \\ \hline
Freqmine & Medium & 512 & 64 & 512 * 64 & 1 & 931933 & 0.44823 & 0.223479 & 3496345 & 40548.71 & 0.34 & 65.8 \\ \hline
Freqmine & Medium & 512 & 256 & 512 * 256 & 1 & 602102 & 0.533319 & 0.265416 & 1821316 & 40220.09 & 0.18 & 71.26 \\ \hline
Freqmine & Medium & 512 & 512 & 512 * 512 & 10 & 517035 & 0.570667 & 0.28436 & 1359778 & 38220.44 & 0.13 & 78.49 \\ \hline
Freqmine & Medium & 512 & 1024 & 512 * 1024 & 1 & 480607 & 0.593974 & 0.296231 & 1102871 & 36834.71 & 0.11 & 83.55 \\ \hline
Freqmine & Medium & 1024 & 16 & 1024 * 16 & 1 & 0 & 0 & 0 & 0 & 0 & 0 & 0 \\ \hline
Freqmine & Medium & 1024 & 64 & 1024 * 64 & 1 & 931933 & 0.44823 & 0.223479 & 3496345 & 40548.71 & 0.34 & 65.8 \\ \hline
Freqmine & Medium & 1024 & 256 & 1024 * 256 & 1 & 602102 & 0.533319 & 0.265416 & 1821316 & 40220.09 & 0.18 & 71.26 \\ \hline
Freqmine & Medium & 1024 & 512 & 1024 * 512 & 1 & 517035 & 0.570667 & 0.28436 & 1359778 & 38220.44 & 0.13 & 78.49 \\ \hline
Freqmine & Medium & 1024 & 1024 & 1024 * 1024 & 10 & 480607 & 0.593974 & 0.296231 & 1102871 & 36834.71 & 0.11 & 83.55 \\ \hline
\end{tabular}
\caption{\texttt{Freqmine} benchmarks for medium simulation on the X86 simulated processor using gem5 simulator.}
\label{tab:medium}
\end{table}
\end{landscape}
\newgeometry{margin=2cm}  
\begin{landscape}
\begin{table}[] 
\scriptsize
\begin{tabular}{|c|c|c|c|c|c|c|c|c|c|c|c|c|}
\hline
\textbf{Benchmark} & \textbf{Size} & \textbf{L1I Size} & \textbf{L1D size} & \textbf{L1I * L1D} & \textbf{\# Runs} & \textbf{Squashed misses} & \textbf{IER} & \textbf{IPC} & \textbf{\# Cache Misses} & \textbf{Avg. Mem. per DRAM} & \textbf{DBU} & \textbf{Hit rate} \\ \hline
Freqmine & Small & 16 & 16 & 16 * 16 & 10 & 0 & 0 & 0 & 0 & 0 & 0 & 0 \\ \hline
Freqmine & Small & 16 & 64 & 16 * 64 & 1 & 932167 & 0.448235 & 0.223482 & 3496299 & 40550.07 & 0.34 & 65.8 \\ \hline
Freqmine & Small & 16 & 256 & 16 *256 & 1 & 601880 & 0.533339 & 0.26546 & 1820694 & 40206.53 & 0.18 & 71.28 \\ \hline
Freqmine & Small & 16 & 512 & 16 *512 & 1 & 517037 & 0.570672 & 0.284355 & 1359632 & 38216.32 & 0.13 & 78.49 \\ \hline
Freqmine & Small & 16 & 1024 & 16 * 1024 & 1 & 480141 & 0.59391 & 0.296251 & 1102909 & 36830.53 & 0.11 & 83.56 \\ \hline
Freqmine & Small & 64 & 16 & 64 * 16 & 1 & 0 & 0 & 0 & 0 & 0 & 0 & 0 \\ \hline
Freqmine & Small & 64 & 64 & 64 * 64 & 10 & 932167 & 0.448235 & 0.223482 & 3496299 & 40550.07 & 0.34 & 65.8 \\ \hline
Freqmine & Small & 64 & 256 & 64 * 256 & 1 & 601880 & 0.533339 & 0.26546 & 1820694 & 40206.53 & 0.18 & 71.28 \\ \hline
Freqmine & Small & 64 & 512 & 64 * 512 & 1 & 517037 & 0.570672 & 0.284355 & 1359632 & 38216.32 & 0.13 & 78.49 \\ \hline
Freqmine & Small & 64 & 1024 & 64 * 1024 & 1 & 480141 & 0.59391 & 0.296251 & 1102909 & 36830.53 & 0.11 & 83.56 \\ \hline
Freqmine & Small & 256 & 16 & 256 * 16 & 1 & 0 & 0 & 0 & 0 & 0 & 0 & 0 \\ \hline
Freqmine & Small & 256 & 64 & 256 * 64 & 1 & 932167 & 0.448235 & 0.223482 & 3496299 & 40550.07 & 0.34 & 65.8 \\ \hline
Freqmine & Small & 256 & 256 & 256 * 256 & 10 & 601880 & 0.533339 & 0.26546 & 1820694 & 40206.53 & 0.18 & 71.28 \\ \hline
Freqmine & Small & 256 & 512 & 256 * 512 & 1 & 517037 & 0.570672 & 0.284355 & 1359632 & 38216.32 & 0.13 & 78.49 \\ \hline
Freqmine & Small & 256 & 1024 & 256 * 1024 & 1 & 480141 & 0.59391 & 0.296251 & 1102909 & 36830.53 & 0.11 & 83.56 \\ \hline
Freqmine & Small & 512 & 16 & 512 * 16 & 1 & 0 & 0 & 0 & 0 & 0 & 0 & 0 \\ \hline
Freqmine & Small & 512 & 64 & 512 * 64 & 1 & 932167 & 0.448235 & 0.223482 & 3496299 & 40550.07 & 0.34 & 65.8 \\ \hline
Freqmine & Small & 512 & 256 & 512 * 256 & 1 & 601880 & 0.533339 & 0.26546 & 1820694 & 40206.53 & 0.18 & 71.28 \\ \hline
Freqmine & Small & 512 & 512 & 512 * 512 & 10 & 517037 & 0.570672 & 0.284355 & 1359632 & 38216.32 & 0.13 & 78.49 \\ \hline
Freqmine & Small & 512 & 1024 & 512 * 1024 & 1 & 480141 & 0.59391 & 0.296251 & 1102909 & 36830.53 & 0.11 & 83.56 \\ \hline
Freqmine & Small & 1024 & 16 & 1024 * 16 & 1 & 0 & 0 & 0 & 0 & 0 & 0 & 0 \\ \hline
Freqmine & Small & 1024 & 64 & 1024 * 64 & 1 & 932167 & 0.448235 & 0.223482 & 3496299 & 40550.07 & 0.34 & 65.8 \\ \hline
Freqmine & Small & 1024 & 256 & 1024 * 256 & 1 & 601880 & 0.533339 & 0.26546 & 1820694 & 40206.53 & 0.18 & 71.28 \\ \hline
Freqmine & Small & 1024 & 512 & 1024 * 512 & 1 & 517037 & 0.570672 & 0.284355 & 1359632 & 38216.32 & 0.13 & 78.49 \\ \hline
Freqmine & Small & 1024 & 1024 & 1024 * 1024 & 10 & 480141 & 0.59391 & 0.296251 & 1102909 & 36830.53 & 0.11 & 83.56 \\ \hline
\end{tabular}
\caption{\texttt{Freqmine} benchmarks for small simulation on the X86 simulated processor using gem5 simulator.}
\label{tab:small}
\end{table}
\end{landscape}
\restoregeometry

\section{Statistical Analysis}

\begin{figure}
\centering

\begin{tikzpicture}[scale=1.25]
    \colorlet{col1}{blue!70}
    \colorlet{col2}{blue!60}
    \colorlet{col3}{blue!50}
    \colorlet{col4}{blue!40}

   \draw[->] (0,-1.25) -- (0,1.5) node [above]
     {};

   \begin{scope}[smooth,draw=gray!20,y=0.3989422804cm]
        \filldraw [fill=col3] plot[id=f1,domain=-3:-2] function {exp(-x*x/2)}
            -- (-2,0) -- (-3,0) -- cycle;
        \filldraw [fill=col2] plot[id=f2,domain=-2:-1] function {exp(-x*x/2)}
            -- (-1,0) -- (-2,0) -- cycle;
        \filldraw [fill=col1] plot[id=f3,domain=-1:0]  function {exp(-x*x/2)}
            -- (0,0)  -- (-1,0) -- cycle;
        \filldraw [fill=col1] plot[id=f4,domain=0:1] function {exp(-x*x/2)}
            -- (1,0)  --  (0,0) -- cycle;
        \filldraw [fill=col2] plot[id=f5,domain=1:2] function {exp(-x*x/2)}
            -- (2,0)  -- (1,0) -- cycle;
        \filldraw [fill=col3] plot[id=f6,domain=2:3] function {exp(-x*x/2)}
            -- (3,0)  -- (2,0) -- cycle;
        \draw[black] plot[id=f7,domain=-4.25:4.25,samples=100]
            function {exp(-x*x/2)};
   \end{scope}
       \draw[->] (-4.25,0) -- (4.25,0) node [right] {$y$};

    \foreach \pos/\label in {-3/$-3\sigma$,-2/$-2\sigma$,-1/$-\sigma$,
            1/$\sigma$,2/$2\sigma$,3/$3\sigma$}
        \draw (\pos,0) -- (\pos,-0.1) (\pos cm,-3ex) node
            [anchor=base,fill=white,inner sep=1pt]  {\label};

    \draw (-0.1,1) -- (.1,1) node [right,fill=white,inner sep=1pt] {$\sigma$};

    \foreach \pos/\percent/\height in {1/34/0.5,2/14/0.25,3/2/0.125,4/0.1/0.1}
    {
      \node[text=col\pos,anchor=base,yshift=2pt,xshift=-0.625cm,
        fill=white,inner sep=1pt] at (\pos,\height) {$\percent\%$};
      \node[text=col\pos,anchor=base,yshift=2pt,xshift=.625cm,
        fill=white,inner sep=1pt]  at (-\pos,\height) {$\percent\%$};
    }
\end{tikzpicture}
\caption{A confidence interval of 95\% is approximately
within 2 standard deviations of the distribution as shown in online  \href{http://www.texample.net/tikz/examples/standard-deviation}{Online Example Standard deviation.}}
\end{figure}
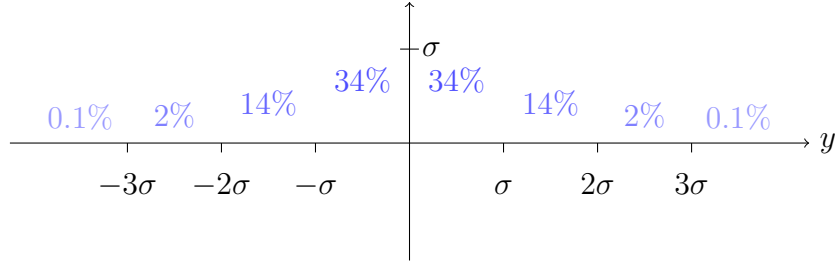
Our statistical significance based computation for a set of observations of any metric is as follows:  
$$\displaystyle
        \frac{1}{\sigma\sqrt{2\pi}}\exp\biggl(\frac{-y^2}{2\sigma^2}\biggr);  \sigma^2 = \sqrt{\sum_{i=1}^n (y_i - \bar{y})^2}.$$

Where $y_i$ is the $i^{th}$ element of the sample, and $y$ is the sample mean, $s$ is the standard deviation of the sample, and $n$ is the number of samples. 

\justify        
We run all experiments 10 times each, but to our surprise we found no changes in the values of the metrics we measure. Thus, we removed the random number generator seed in the \verb+random.cc+ file and recompiled the entire system, and perform totally random (pseudo-random in theory!) experiments. We still see no changes in the results spanning all the 10 experiments. The deviations (which can be observed in the three tables) are due to simulation scale (small vs. medium vs. large). Representative examples on Hit Rate, and IPC are shown below. The minor variations lead us to conclude that with 95\% confidence, true distribution mean is close to our observed mean. Since other metrics also observe minor variations, we argue the true mean lies close to observed values allowing us to skip CI computations for the rest of them.

\underline{Example 1}: Observed Hit Rates for \texttt{16*256} setup are: [71.28, 71.26, 71.28]. Since each value is across 10 experimental runs, $n = 30$. From the given formula, we obtain the true range \textbf{71.3 $\pm$ 0.013} with a 95\% CI.

\underline{Example 2}: Observed IPC for \texttt{16*64} setup are: [0.223479, 0.223482, 0.223482]. Since each value is across 10 experimental runs, $n = 30$. From the given formula, we obtain the true range \textbf{0.223 $\pm$ 1.96e-6} with a 95\% CI.

These calculations can be verified using the online CI calculator\footnote{\href{https://www.mathsisfun.com/data/confidence-interval-calculator.html}{Online Confidence Interval Calculator}}.

\section{Conclusion}
We perform a myriad of simulations (\textbf{\textit{\textbf{> 500}}}) on three simulation scales - small, medium, and large to draw the following inferences and observations:
\begin{itemize}
    \item Contrary to our expectations, changing L1I cache size has no effect whatsoever on any relevant metrics for \texttt{Freqmine} simulation at any scale. We hypothesize that this is the case because \texttt{Freqmine} is a data-intensive application. Our experiments covered an ultimately wide range of I-cache sizes from 1 KB to 1024 KB allowing us to make us confident inferences from our observations.
    
    \item Size of L1D cache is vital to performance of \texttt{Freqmine} simulations in X86 system and is directly proportional to Hit Rate which aligns with our expectations. Since, this benchmark shows high spatial locality for data with quick temporal reuse of instructions (i.e. as highly parallel loops are common in data-mining applications like \texttt{Freqmine}), increase in L1D also corresponds to a higher IPC.
    
    \item We observed that if the size of L1D cache is lower than a certain threshold (64 KB in our case) the X86 system crashes during simulation. Hence, we believe that the importance of data cache size trumps the importance of instruction cache size for our use case i.e. data-mining application such as Freqmine.
    
    \item Even after going through extreme lengths to introduce randomness, we observed that the performance stays constant for multiple runs when all parameters are fixed. The only variation we observe is when the simulation scale is changed (small vs. medium vs. large) which allowed us to compute confidence interval based statistics.
    
\end{itemize}

\section{Acknowledgments} We would like to thank Prof. Dan Marinescu and Dr. Debashri Roy for their timely, and helpful insights not only for this project but also on overall computer architecture theory.

\end{document}